# Ultrasound-based detection and malignancy prediction of breast lesions eligible for biopsy: A multi-center clinical-scenario study using nomograms, large language models, and radiologist evaluation

Ali Abbasian Ardakani[1*], PhD; Afshin Mohammadi[2], MD; Taha Yusuf Kuzan[3], MD; Beyza Nur Kuzan[4], MD; Hamid Khorshidi[5], MSc; Ashkan Ghorbani[1] BSc; Alisa Mohebbi[6,7], MD; Fariborz Faeghi[1], PhD; Sepideh Hatamikia[8,9], PhD; U Rajendra Acharya[10,11], PhD, DEng, DSc

[1] *Department of Radiology Technology, School of Allied Medical Sciences, Shahid Beheshti University of Medical Sciences, Tehran, Iran*
[2] *Department of Radiology, Faculty of Medicine, Urmia University of Medical Science, Urmia, Iran*
[3] *Department of Radiology Sancaktepe Şehit Prof. Dr. İlhan Varank Training and Research Hospital, University of Health Sciences, Istanbul, Turkey*
[4] *Kartal Dr. Lütfi Kırdar City Hospital, Istanbul, Turkey*
[5] *Department of Information Engineering, University of Padova, Padova, Italy*
[6] *Universal Scientific Education and Research Network (USERN), Tehran, Iran*
[7] *School of Medicine, Tehran University of Medical Sciences, Tehran, Iran*
[8] *Clinical AI-Research in Omics and Medical Data Science (CAROM) group, Department of Medicine, Faculty of Medicine and Dentistry, Danube Private University, Krems an der Donau, Austria*
[9] *Austrian Center for Medical Innovation and Technology (ACMIT), Wiener Neustadt, Austria*
[10] *School of Mathematics, Physics and Computing, University of Southern Queensland, Springfield, Queensland, Australia*
[11] *Centre for Health Research, University of Southern Queensland, Springfield, Queensland, Australia*

**∗ Corresponding author:**
**Dr. Ali Abbasian Ardakani**
**E-mail: Ardakani@sbmu.ac.ir, A.ardekani@live.com. ORCID: 0000-0001-7536-0973**

# Ultrasound-based detection and malignancy prediction of breast lesions eligible for biopsy: A multi-center clinical-scenario study using nomograms, large language models, and radiologist evaluation

## Take-home Message

- The proposed AI model significantly outperformed radiologists and LLMs, achieving accuracies of 83.0% for biopsy recommendation and 83.8% for malignancy prediction, respectively.
- The proposed AI model adheres to clinical scenarios for biopsy recommendations and malignancy risk estimation, offering the potential to reduce unnecessary interventions and optimize clinical decision-making in breast imaging.

**Abstract**

**Rationale and Objectives:** To develop and externally validate ultrasound nomograms combining BI-RADS features and quantitative morphometric characteristics, and to compare their performance with expert radiologists and large language models in biopsy recommendation and malignancy prediction for breast lesions.

**Methods:** In this multi-center, multi-national study, 1,747 women with breast lesions underwent ultrasound across three centers in Iran and Turkey. A total of 10 BIRADS and 26 morphological features were extracted from each lesion. Three nomograms based on BI-RADS, morphometric, and both feature sets were constructed. Three radiologists (one senior, two general) and two ChatGPTs including ChatGPT-o3 and o4-mini-high interpreted de-identified breast lesion images. Diagnostic performance for biopsy recommendation and malignancy prediction was assessed across all cohorts.

**Results:** According to the pooled results, although the difference between the fused nomogram and the BI-RADS version was not statistically significant, the fused version consistently outperformed all models in biopsy recommendation and malignancy prediction (AUCs of 0.901 and 0.853, respectively) compared to BI-RADS nomogram (AUCs of 0.898 and 0.834), morphometric nomogram (AUCs of 0.825 and 0.708), radiologist1 (AUCs of 0.820 and 0.729), radiologist2 (AUCs of 0.605 and 0.719), radiologist3 (AUCs of 0.728 and 0.699), ChatGPT-o3 (AUCs of 0.729 and 0.689), and o4-mini-high (AUCs of 0.713 and 0.695).

**Conclusions:** The proposed BI-RADS–morphometric nomogram outperforms standalone nomogram models, LLMs, and radiologists in guiding biopsy decisions and predicting malignancy. The proposed novel fused nomogram has the potential to reduce unnecessary biopsies and enhance personalized decision-making in breast imaging.

## 1. Introduction

The Breast Imaging Reporting and Data System (BI-RADS), developed by the American College of Radiology (ACR), has provided standardized terminology and classification systems for mammography, ultrasound, and magnetic resonance imaging of the breast (1-3). However, despite these standardized guidelines, significant challenges persist in breast imaging interpretation, including substantial interobserver variability among radiologists and the inherent complexity of distinguishing benign from malignant lesions (4, 5).

The emergence of artificial intelligence (AI) technologies has opened new avenues for improving diagnostic accuracy in breast imaging. In this regard, Large Language Models (LLMs), particularly ChatGPT variants, have demonstrated remarkable potential in medical imaging interpretation and clinical decision-making (6). Recent investigations have demonstrated that ChatGPT-4 can achieve comparable performance to experienced radiologists in BI-RADS classification and malignancy prediction from raw reports. Notably, LLMs have shown particular promise in reducing diagnostic variability among less experienced radiologists while maintaining consistency with senior clinicians (7).

Simultaneously, advances in quantitative imaging have enabled the extraction of high-dimensional morphological features from breast ultrasound images. These quantitative characteristics, including lesion shape, margin irregularity, echogenicity patterns, and textural heterogeneity, have shown significant potential for automated lesion classification. Studies have demonstrated that quantitative morphological features can achieve AUC values ranging from 0.74 to 0.94 in distinguishing benign from malignant breast lesions (8-10).

In addition to LLMs and morphological features, the development of nomograms using imaging features has emerged as a powerful approach for risk stratification and clinical decision-making. These predictive models integrate multiple diagnostic parameters to provide individualized malignancy probabilities, potentially reducing unnecessary biopsies while maintaining high sensitivity for cancer detection (11-14). However, the clinical validation of such models across diverse populations and healthcare settings remains challenging, with many studies limited by single-center designs and inadequate external validation.

Despite these technological advances, comprehensive comparative studies evaluating the performance of LLMs, morphologic-based models, traditional and nomographic BI-RADS assessment, and radiologist interpretation across different experience levels remain critically limited. Importantly, most existing studies investigating LLMs in medical imaging have primarily utilized pre-existing radiologist reports rather than direct interpretation of imaging data. This fundamental limitation has prevented a true assessment of LLMs' independent diagnostic capabilities, as they have essentially been functioning as sophisticated text processors rather than genuine imaging interpreters. Furthermore, the integration of these approaches into unified diagnostic frameworks that could potentially optimize both biopsy decision-making and malignancy prediction has not been thoroughly investigated across diverse international populations. This study aims to address these critical gaps by developing and externally validating integrated nomogram models that combine BI-RADS features with quantitative morphological characteristics, while simultaneously evaluating the performance of state-of-the-art LLMs (ChatGPT-o3 and o4-mini) in direct ultrasound image interpretation against radiologists of varying experience levels. Our comprehensive multi-center and multi-national validation represents a paradigm shift from traditional radiology and report-based LLM assessments to genuine imaging analysis, providing crucial insights into the true generalizability and clinical utility of these emerging diagnostic technologies in breast cancer.

## 2. Methods

### 2.1. Patient Selection & Study Design

This multi-center, multi-national retrospective study was conducted across three specialized breast imaging centers to develop, and to internally and externally validate models assessing the performance of LLMs, a selected BI-RADS features-based nomogram, a morphometric nomogram, a BI-RADS-morphometric nomogram, and three radiologists with different levels of experience in biopsy recommendation and malignancy prediction. Institutional Review Board (IRB) approval was obtained from all participating centers (IRB #2025/153 and #IR.SBMU.RETECH.REC.1404.338). All procedures were performed in accordance with the ethical standards of the institutional and national research committees. The study was designed in accordance with established guidelines for multi-center validation studies in AI applications for medical imaging and nomogram development in breast cancer research.

The study population consisted of consecutive patients presenting with breast lesions requiring ultrasound evaluation between September 2023 and June 2025. Inclusion criteria were rigorously standardized across all participating centers and encompassed: (A) patients aged >18 years with breast lesions detected on ultrasound examination with BI-RADS 2 and more, (B) a definitive histopathological diagnosis obtained through core needle biopsy or fine needle aspiration; for lesions classified as BI-RADS 2 and 3, follow-up was conducted in accordance with ACR guidelines, and (C) ultrasound examination performed within four days prior to tissue sampling. Patients were excluded based on predefined criteria to maintain data integrity and minimize confounding variables: (A) patients who declined to provide informed consent for research participation or data utilization; (B) individuals who had received any form of breast cancer treatment including chemotherapy, radiation therapy, surgical intervention, or hormonal therapy prior to initial imaging, which could alter tissue characteristics and imaging appearance; (C) history of breast reconstructive surgery; (D) presence of breast implants, which alter acoustic properties and imaging characteristics; (E) inconclusive or borderline pathological results that could not be definitively classified as benign or malignant according to established histopathological criteria; (F) recurrent mastitis or active inflammatory conditions, particularly idiopathic granulomatous mastitis, which could mimic malignant imaging features, and (G) technical limitations preventing adequate ultrasound visualization including severe obesity with inadequate acoustic penetration (BMI>40 kg/m$^2$), chest wall deformities, or extensive scarring from prior procedures.

To ensure consistent and unbiased analysis, only the largest lesion per patient, as identified on ultrasound examination, was evaluated in the study, following established protocols for multi-lesion breast imaging research. This approach was implemented to avoid potential clustering effects within patients and to maintain independence of observations for statistical analysis. In cases where multiple lesions of similar size were present, the most suspicious lesion based on BI-RADS features was selected for analysis. Lesion measurements were performed in two orthogonal planes using standardized measurement techniques.

A total of 1,831 patients were initially assessed across all participating centers, representing consecutive cases meeting broad inclusion criteria. Following systematic application of exclusion criteria, 84 patients (4.5%) were excluded. This resulted in 1,747 patients for the final analysis, distributed as follows: 1,127 patients from Iranian center 1, 345 patients from Iranian Center 2, and 275 patients from Turkish Center 3 (Figure 1).

**2.2. Pathological Classification**

Histopathological diagnosis served as the definitive gold standard for BI-RADS 4 and 5 lesions included in the study. Tissue sampling was performed using standardized protocols across all participating centers. All tissue samples were processed using standardized histological techniques with formalin fixation and paraffin embedding, followed by hematoxylin and eosin staining for routine morphological assessment.

Pathological classification followed the International Classification of Diseases, 10th Revision (ICD-10) coding system with rigorous adherence to established diagnostic criteria. Malignant breast lesions were defined as those histopathologically classifiable under ICD-10 C50 codes, encompassing invasive ductal carcinoma, invasive lobular carcinoma, ductal carcinoma in situ, lobular carcinoma in situ, mixed invasive carcinomas, and rare histological subtypes such as mucinous, papillary, medullary, and metaplastic carcinomas. Benign lesions were classified under ICD-10-CM D24.1 codes, encompassing fibroadenomas, phyllodes tumors, papillomas, sclerosing adenosis, radial scars, lipomas, hamartomas, and other benign breast neoplasms.

Histopathological evaluation was conducted by board-certified pathologists with specialized expertise in breast pathology at each participating center. For quality assurance and standardization, challenging cases were reviewed through multidisciplinary conferences with additional consultation from external pathology experts as needed.

### 2.3. Ultrasound Image Acquisition & Protocol

Standardized imaging protocols were implemented across all centers to minimize technical variability and ensure reproducible image quality. Each examination began with a comprehensive bilateral breast ultrasound performed by board-certified radiologists or experienced sonographers under the direct supervision of a radiologist. The imaging protocol included systematic evaluation of both breasts using multiple scanning approaches, including radial, anti-radial, and cross-sectional techniques to ensure complete tissue coverage. Patient positioning followed a uniform protocol: supine oblique for the lateral quadrants, supine with the ipsilateral arm raised for medial and retroareolar regions, and contralateral decubitus for deep posterior lesions. Gentle compression techniques were applied to optimize lesion conspicuity while avoiding tissue deformation that could distort morphometric analysis. Specific attention was directed toward the detection and characterization of focal lesions, with detailed documentation of lesion location using the clock-face method and the measurement of distance from the nipple.

Frequency selection was adjusted between 7-15 MHz based on patient habitus and lesion depth, a with preference for higher frequencies when adequate penetration was achieved. Image storage and management adhered to standardized digital imaging protocols with lossless compression formats to preserve image quality for subsequent analysis. All images were archived using institutional Picture Archiving and Communication Systems (PACS), which included detailed metadata including acquisition parameters, patient positioning, and lesion identification markers. Quality control measures included immediate image review for technical adequacy, standardized annotation systems for lesion identification, and systematic verification of image-to-pathology correlation for all cases.

### 2.4. LLM-Based Interpretation Workflow

Two state-of-the-art ChatGPT variants were selected for evaluation: ChatGPT-4o-mini-high and ChatGPT-o3. The methodology for ChatGPT evaluation was designed to simulate real-world clinical decision-making scenarios while maintaining standardized assessment protocols. To ensure consistent input data, the Region of Interest (ROI) containing the lesion was cropped from the original ultrasound image, matching the view provided to radiologists, and exported as a high-resolution JPG file at 300 dpi. This cropping step removed peripheral annotations and scanner metadata to prevent data leakage. The models were provided with a standardized, academically structured prompt (detailed in Supplementary Table S1) designed to elicit comprehensive diagnostic assessments by assigning the model an expert persona. A zero-shot prompting approach was used without any fine-tuning. The prompting strategy incorporated multiple assessment tasks including BI-RADS feature description using established ultrasound terminology, BI-RADS classification assignment with supporting rationale, and whether the lesion is malignant or benign if biopsied. The models were instructed to analyze each image independently without access to clinical history, patient demographics, histopathology, or previous imaging studies, thereby replicating the blinded interpretation conditions applied to human radiologists. To quantify runtime performance, each model's response latency was recorded. ChatGPT-4o-mini-high generated outputs within 10-30 seconds per image, while ChatGPT-o3 required 30-120 seconds.

### 2.5. Radiologist Evaluation & BI-RADS Nomogram Assessment

Three radiologists participated in the image interpretation process, representing a spectrum of clinical experience and specialization levels. The senior radiologist possessed 24 years of comprehensive experience in breast imaging. Two general radiologists from the Turkish institution participated, each with 5 and 7 years of experience in general radiology with regular breast imaging responsibilities. For BI-RADS feature reporting, we selected the most commonly reported item. In cases where no conclusion was reached, consensus was achieved through direct discussion.

All interpretations adhered strictly to the Fifth Edition of the ACR BI-RADS Ultrasound Lexicon. Radiologists systematically documented each lesion’s tissue composition, shape, orientation, margin characteristics, echo pattern, posterior acoustic features (no posterior features; enhancement; shadowing), calcification status, architectural distortion, presence of clustered microcysts, and complicated cyst attributes. Assignment of the appropriate BI-RADS category ranging from 2 to 5 was performed based on the aggregate imaging findings.

Radiologists reviewed images using high-resolution PACS workstations with standardized viewing conditions, including controlled ambient lighting, calibrated monitors, and optimized window/level settings. Each case was interpreted independently without access to clinical history, patient demographics, prior imaging studies, pathology, or other radiologists' interpretations to ensure an unbiased assessment. The interpretation protocol allowed unlimited review time for each case.

For each lesion, the three radiologists first indicated whether tissue sampling would be recommended (i.e., whether the lesion met BI-RADS category 4 or 5 criteria) as the first outcome of the study. Subsequently, radiologists recorded their personal impression of the lesion’s benign versus malignant nature based on imaging criteria and their clinical experience as the second outcome of the study, if it was biopsy candid. The decision to perform a biopsy (i.e., biopsy candidate) was assessed as a reference test for evaluation using a consensus-based approach that reflected real-world multidisciplinary practice patterns. Three radiologists independently reviewed each case and provided biopsy recommendations based on their BI-RADS assessments and clinical judgment (BI-RADS 4 and above). Cases were classified as biopsy candidates if two or three radiologists recommended tissue sampling, while cases with zero or one biopsy recommendation were considered non-candidates. This consensus approach was designed to reflect typical multidisciplinary team decision-making processes while providing a robust reference standard for evaluating automated decision-support systems.

To develop the BI-RADS nomogram using the rms package in R, all imaging features were first subjected to univariable logistic regression analysis. Only features achieving a significance threshold of $p<0.05$ in univariable testing were carried forward into multivariable modeling. Multivariable logistic regression then identified the independent predictors of biopsy recommendation and malignancy; again, only variables with $p<0.05$ were retained in the final regression equation to assess both outcomes.

### 2.6. Feature Extraction for Morphological Nomogram

The morphometric analysis framework was designed to extract quantitative shape characteristics that could provide objective, reproducible measurements of lesion properties while remaining independent of operator variability and technical imaging parameters. Feature selection employed a rigorous three-step process designed to identify the most discriminative and robust characteristics while minimizing redundancy and overfitting.

In the first step, lesion segmentation was independently performed by two radiologists to ensure accurate boundary delineation. For quality assurance, inter-observer agreement was assessed using intraclass correlation coefficients (ICC). Only features demonstrating ICC values greater than 0.85 were accepted for subsequent analysis to ensure reproducibility across different operators. From each ROI, a set of 26 morphological features was extracted, encompassing geometric characteristics that reflected fundamental shape properties relevant to malignancy assessment. These features included area and perimeter measurements providing basic size quantification, shape descriptors such as circularity, roundness, and compactness that captured lesion regularity, margin characteristics including convexity and concavity measurements that reflected boundary smoothness, aspect ratios and elongation parameters that described lesion orientation and dimensional relationships, complexity measures such as extent and orientation angle that characterized irregular growth patterns, and specialized geometric features including rectangularity and inscribed/circumscribed circle ratios that provided additional shape discrimination capability. The definition of all morphological features is provided in Figure 2.

The second step involved collinearity assessment using the Pearson correlation coefficient. Features with Pearson correlation coefficient values above 0.85 were excluded from subsequent analysis to facilitate dimensionality reduction. From the features with high collinearity, the one with the lower univariable AUC was excluded. As the last step, the Least Absolute Shrinkage and Selection Operator (LASSO) regression was applied as the final feature selection technique, providing automated variable selection with built-in regularization to prevent overfitting. Then, the morphological nomogram was constructed using the same methodological framework as the BI-RADS nomogram. Additionally, a fused nomogram integrating both selected BI-RADS and morphological features was created. The three-step quantitative feature selection pipeline was used only for the construction of the morphometric and fused nomograms. The radiologists and ChatGPT models were not restricted to these selected features; rather, they evaluated the full ultrasound images using standard qualitative BI-RADS descriptors and expert pattern recognition to intentionally contrast the diagnostic performance of mathematically selected quantitative features against the holistic visual interpretation performed by human experts and LLMs.

### 2.7. Statistical Analysis & Model Performance Evaluation

All statistical analyses were performed using R statistical software, version 4.3.0, with appropriate packages for specialized analyses. Statistical significance was set at $p<0.05$. All confidence intervals were calculated at the 95% level. Categorical variables were described using frequencies and percentages, with 95% confidence intervals.

The Matthews Correlation Coefficient (MCC) was utilized as the primary correlation measure for binary classification performance, offering significant advantages over traditional kappa statistics commonly employed in medical research. The MCC ranges from -1 to +1 with clear interpretability: values near +1 indicate perfect positive correlation, values near 0 indicate random performance, and values near -1 indicate systematic misclassification.

Receiver operating characteristic (ROC) analysis was performed for all diagnostic methods to evaluate discrimination capability across different decision thresholds. Area under the ROC curve (AUC) values were calculated with 95% confidence intervals. A comparative analysis between different diagnostic methods employed DeLong's test for comparing ROC curves, providing a statistical assessment of performance differences between nomograms, radiologist interpretations, and ChatGPT evaluations.

## 3. Results

### 3.1. Patient Characteristics

The present multi-center, multinational retrospective study comprised a total of 1,747 patients who met the inclusion criteria after rigorous application of predefined exclusion factors. Among these, 1,127 patients were recruited from the first Iranian center (714 biopsy candidates), of which 958 constituted the training cohort (607 biopsy candidates) and 169 formed the internal validation cohort (107 biopsy candidates). An external validation cohort included 345 patients from an additional Iranian center (207 biopsy candidates) and 275 patients from the Turkish center (204 biopsy candidates).

### 3.2. Performance in Biopsy Recommendation

In the training dataset, both univariate and multivariate logistic regression identified several BI-RADS ultrasound features that were significantly associated with biopsy recommendation (Table 1). In multivariate analysis, irregular lesion shape, nonparallel orientation, circumscribed margin, posterior enhancement, and in-mass calcifications remained independent predictors of biopsy candidacy. On the other hand, the three-step quantitative feature selection method identified five morphological features for developing nomograms: Convexity1, Convexity4, Ellipticity, Aspect_ratio, and Concavity (Figure 3).

The primary aim of the first analysis was to evaluate the ability of the BI-RADS nomogram, the morphometric nomogram, the fused nomogram, three radiologists of varying experience levels, and two ChatGPT variants to correctly identify lesions warranting biopsy (BI-RADS 4 or 5 criteria).

In the internal validation cohort, the morphologic nomogram exhibited the highest accuracy among the automated models, correctly classifying 80.5% of lesions for biopsy recommendation. The BI-RADS nomogram yielded an accuracy of 76.3%, matching that of the fused nomogram, which also achieved 76.3% accuracy. Among radiologists, the senior radiologist demonstrated 76.9% accuracy, narrowly surpassing the accuracy of the nomograms, whereas the general radiologists achieved 74.0% and 71.6% accuracy, respectively. The ChatGPT o3 and o4-mini-high models showed lower accuracy in this cohort, with 62.1% and 61.5%, respectively (Table S2).

External validation in the Iranian cohort revealed superior performance of both the BI-RADS and fused nomograms, with accuracies of 89.6% and 90.1%, respectively, substantially exceeding the morphologic nomogram's 77.4% accuracy. The senior radiologist achieved an accuracy rate of 88.7%, while the general radiologists recorded rates of 60.0% and 69.3%. ChatGPT models displayed improved performance in this external cohort, with o3 achieving 81.4% and o4-mini-high 81.4% accuracy (Table S3).

In the Turkish external cohort, both the BI-RADS and fused nomograms again led, each attaining 78.2% accuracy compared to the morphologic nomogram's 69.4% accuracy. The senior radiologist achieved 77.4% accuracy; the general radiologists performed poorly, with one achieving 54.9% accuracy and the other 77.8%. ChatGPT o3 and o4-mini-high recorded 69.1% and 64.4% accuracy, respectively (Table S4).

Aggregating results across all cohorts reinforced the consistent superiority of the BI-RADS and fused nomograms, which achieved accuracies of 82.7% and 83.0%, respectively, compared to the morphologic nomogram's 75.3% accuracy. The senior radiologist's accuracy was 82.2%, whereas general radiologists recorded 61.2% and 72.5% accuracy. ChatGPT o3 achieved 73.0% accuracy and ChatGPT o4-mini-high 71.2% accuracy (Table 2).

Comparative statistical analysis confirmed that the BI-RADS and fused nomograms significantly outperformed the morphologic nomogram, three radiologists and ChatGPT models across all cohorts. The morphologic nomogram outperformed general radiologists but fell short of the performance of senior radiologists. ChatGPT performance, while improving in external cohorts, remained below that of both nomograms and the senior radiologist (Table S5) (Figure 4).

Figure 7A depicts the nomogram developed for biopsy recommendation, illustrating the weighted contributions of each significant BI-RADS and morphological feature. Each predictor's score aligns with its regression coefficient in the multivariate model, enabling individualized risk calculation for biopsy candidacy.

### 3.3. Performance in Malignancy Diagnosis

Univariate and multivariate analyses of BI-RADS features in the training cohort revealed parallel predictors of malignancy (Table 3). In multivariate logistic regression, parallel orientation (odds ratio, 0.44), circumscribed margin (odds ratio, 1.45), and in-mass calcifications (odds ratio, 0.59) emerged as independent predictors of malignancy. Moreover, the proposed quantitative feature selection method identified four morphological features for developing nomograms: Circularity2, Extent, Humoment2, and Ellipticity (Figure 5).

In the internal validation cohort, the fused nomogram achieved the highest accuracy of 82.2% in malignancy diagnosis, followed by the BI-RADS nomogram at 80.4% accuracy and the morphologic nomogram at 75.7% accuracy. The senior radiologist attained 75.7% accuracy, while the general radiologists recorded 53.3% and 56.1% accuracy. ChatGPT o3 and o4-mini-high achieved 63.5% and 60.7% accuracy, respectively (Table S6).

External validation within the second Iranian cohort demonstrated notable improvements for both fused and BI-RADS nomograms, which achieved accuracies of 89.9% and 89.3%, respectively. The morphologic nomogram achieved an accuracy of 78.7%. The senior radiologist's accuracy was 89.8%, closely aligned with the nomograms. General radiologists attained 72.5% and 55.6% accuracy, while ChatGPT o3 and o4-mini-high recorded 83.6% and 83.1% accuracy, respectively (Table S7).

In the Turkish cohort, the fused nomogram again led with 78.4% accuracy, while the BI-RADS nomogram achieved 75.4% accuracy and the morphologic nomogram 71.5% accuracy. The senior radiologist matched closely with 77.0% accuracy. General radiologists achieved 74.0% and 78.4% accuracy. ChatGPT o3 and o4-mini-high attained 69.6% and 61.3% accuracy, respectively (Table S8).

Pooling all cohorts highlighted the consistent strength of the fused nomogram, which achieved an accuracy of 83.8%, surpassing both the BI-RADS nomogram (82.0%) and the morphologic nomogram (75.2%). The senior radiologist achieved an accuracy rate of 81.8%, while the general radiologists recorded rates of 69.1% and 64.7%. ChatGPT o3 attained 73.9% accuracy and ChatGPT o4-mini-high attained 69.9% accuracy (Table 4).

Statistical comparisons revealed that the BIRADS and fused nomograms significantly outperformed the morphometric nomogram, as well as the results of three radiologists and all ChatGPT models, in malignancy diagnosis. The senior radiologist's performance was comparable to that of general radiologists and ChatGPT models but inferior to the BI-RADS and fused nomograms. ChatGPT models, while demonstrating respectable performance in the first external cohort, consistently underperformed relative to nomograms and the senior radiologist (Table S9) (Figure 6).

Figure 7b illustrates the malignancy diagnosis nomogram, highlighting the combined impact of BI-RADS features and selected morphometric characteristics. The nomogram provides a clear graphical tool for clinicians to estimate the probability of malignancy in lesions deemed biopsy candidates. Clinical examples of the provided results are also represented in Figure 8.

## 4. Discussion

The present multi-center, multi-national study provides compelling evidence that integrated nomogram models combining BI-RADS ultrasound features and quantitative morphometric characteristics yield superior diagnostic performance in both biopsy recommendation and malignancy prediction for breast lesions, consistently outperforming standalone morphometric models, LLMs, and senior and general radiologists. The fused nomogram demonstrated the highest accuracy across all cohorts for both biopsy candidacy (83.0%) and malignancy diagnosis (83.8%), significantly exceeding the performance of a senior radiologist with 24 years of experience (82.2% and 81.8%, respectively). These findings underscore the clinical potential of hybrid analytic frameworks that synthesize established imaging lexicons with high-dimensional morphological data to inform individualized patient management.

This study has several strengths. First, the multi-center, multi-national design ensured representation of diverse Middle Eastern and Mediterranean populations, encompassing variations in breast cancer prevalence, healthcare infrastructure, and imaging equipment specifications. By including distinct external cohorts from Turkish and additional Iranian centers, we validated the generalizability of our nomograms across different ultrasound platforms (Supersonic Imagine AixPlorer vs. Toshiba Aplio and Samsung RS85) and operator protocols, addressing a limitation of many prior single-center investigations. Second, the rigorous feature selection pipeline employed for the morphological features, incorporating inter-observer ICC filtering, Pearson collinearity assessment, and LASSO regularization, ensured robust identification of reproducible and non-redundant shape descriptors, thereby mitigating overfitting while maintaining parsimony. Third, the integration of BI-RADS features through univariable and multivariable logistic regression into a user-friendly nomogram capitalizes on established clinical risk factors, ensuring interpretability and fostering radiologist trust, which remains a key barrier to the adoption of AI in practice. Finally, by directly comparing the performance of state-of-the-art ChatGPT variants (o3 and o4-mini-high) in interpreting ultrasound images with that of three radiologists of varying experience levels, this work delineates the current limitations of LLMs as genuine imaging interpreters, rather than sophisticated text processors.

Previous investigations of ultrasound-based radiomics have reported AUCs ranging from 0.78 to 0.94 for distinguishing benign from malignant lesions (15), consistent with the performance of our morphologic nomogram (AUC 0.825 for biopsy recommendation and 0.708 for malignancy diagnosis). However, many of those studies were limited by single-center designs, lack of external validation, and reliance on abstract texture features that hinder clinical interpretability. By contrast, our morphological features, such as circularity, concavity, and convexity, are intuitively linked to lesion irregularity and growth patterns, facilitating radiologist acceptance and potential integration into PACS workflows.

Nomograms based solely on BI-RADS features have demonstrated promising discrimination in prior reports using required prospective, multi-center validation (16, 17). Our BI-RADS nomogram achieved AUCs of 0.898 for biopsy recommendation and 0.834 for malignancy diagnosis across all cohorts, matching or surpassing published models (14). Importantly, the fused nomogram significantly improved both sensitivity and specificity by harnessing complementary information from morphological data, echoing findings in invasive status prediction studies that combined radiomics with clinical variables (18). The utility of these morphological features such as convexity and circularity likely lies in their ability to objectively quantify subtle margin irregularities that are subject to high inter-observer variability in visual assessment. For instance, distinguishing between a benign macrolobulated margin and a suspicious microlobulated margin can be challenging for radiologists. The quantitative metrics provide a mathematical value in these ambiguous cases, helping to correctly reclassify benign lesions with deceptive irregular contours that might otherwise be biopsied based on subjective impression alone (Fig. 8e-f). This objective characterization of lesion geometry thus serves to reduce false positives and false negatives without compromising the high specificity and sensitivity required for malignancy detection.

It is noteworthy that out of the eight diagnostic modalities evaluated in this study (including three radiologists, three nomograms and two advanced LLMs) the fused and BI-RADS nomograms consistently ranked as the top two performers. While the improvement of the fused model over the pure BI-RADS model appears modest (e.g., AUC improvement from 0.898 to 0.901), this must be interpreted in the context of a performance ceiling. The BI-RADS nomogram already operates at a high diagnostic threshold, making further aggregate gains difficult to achieve. The value of the fused approach lies in its ability to integrate objective, non-visual quantitative metrics that may resolve complicated cases (Fig. 8e-f, as discussed earlier) where qualitative assessment is equivocal, thus providing a marginal but clinically relevant layer of robustness without sacrificing the interpretability of the standard BI-RADS lexicon.

The introduction of LLMs into imaging interpretation is an emerging frontier. ChatGPT-4 has shown impressive accuracy (98.4%) in recommending appropriate imaging modalities for breast cancer screening and moderate performance (77.7%) for breast pain indications (19). Besides, the agreement between ChatGPT-4o and two expert radiologists in assigning BI-RADS categories to breast lesions on ultrasound images was moderate to substantial (20). In addition, Miaojiao et al. indicated that the performance of ChatGPT-4 was similar to that of two senior radiologists and statistically outperformed two junior radiologists in classifying benign and malignant breast lesions on ultrasound images (7). Motivated by this emerging evidence, we evaluated the model's performance specifically on breast ultrasound within a realistic clinical scenario, which allowed us to assess not only the model's image-interpretation ability but also its behavior in a realistic diagnostic workflow. Our results demonstrated that the diagnostic performance was comparable to that of one general radiologist and significantly higher than that of another in identifying breast lesions that were candidates for biopsy, and its ability to distinguish benign from malignant biopsy-proven lesions was similar to both general and senior radiologists (Table S5 and S9). However, their performance remained inferior to the BI-RADS and fused nomograms. This performance gap is clinically consequential in breast imaging, where small degradations in specificity can translate into substantially higher benign biopsy rates, and small degradations in sensitivity risk delayed cancer diagnosis (21). These results indicate that, despite the known limitations of current vision encoders, LLM-based tools already possess meaningful capacity for image analysis. By documenting both the strengths and the limitations, our study offers a realistic benchmark that can guide future model refinement and help define the role of multimodal LLMs in breast ultrasound interpretation. However, the findings of the present study emphasize that current multimodal LLMs are not yet clinically ready for autonomous ultrasound-based decision-making.

The robust performance of the fused nomogram has significant clinical ramifications. First, by accurately distinguishing lesions requiring biopsy (BI-RADS 4-5) from those amenable to surveillance, the nomogram can reduce unnecessary interventions and associated morbidity, anxiety, and healthcare costs. With specificity exceeding 72% across cohorts and AUCs above 0.89, implementation of the nomogram could substantially lower benign biopsy rates while maintaining high sensitivity for cancer detection, aligning with the goals of clinical decision support in breast imaging. Second, the malignancy-prediction nomogram facilitates personalized risk stratification among biopsy candidates, potentially guiding preoperative planning and patient counseling. The ability to estimate individualized malignancy probability with high accuracy (83.8%) empowers clinicians to tailor management strategies, such as performing a biopsy for high-risk cases or adjunctive imaging for intermediate-probability lesions. Finally, the nomogram's interpretability and reliance on BI-RADS descriptors and intuitive morphometric metrics promote integration into routine practice and enhance radiologists confidence, addressing common barriers to AI adoption in imaging (22).

Taken together, the main practical value of the proposed model is that it can provide clear, usable outputs that fit into routine clinical practice. It can be used as an interpretable decision-support tool at the point of care. In a prospective workflow, radiologists can input routine BI-RADS descriptors already reported in standard ultrasound documentation and a lesion contour. The model can then provide two clinically useful outputs: (i) the probability that a lesion meets biopsy candidacy and (ii) the probability of malignancy among biopsy-eligible lesions. In practice, this can serve as a “second reader”, especially in challenging situations such as BI-RADS 3 versus 4A decisions or borderline 4A lesions. The nomogram may help clinicians communicate risk more clearly and make decisions with patients, for example, by monitoring lower-risk lesions and prioritizing faster workup for higher-risk lesions. It can facilitate risk-stratified enrollment in prospective validation cohorts or clinical trials, particularly by targeting intermediate-risk lesions where uncertainty is greatest and adjunctive strategies may offer the most benefit.

Future research should focus on integrating automatic segmentation pipelines with the nomogram framework to enable real-time risk estimation at the point of care. Additionally, combining text-based clinical information (e.g., patient history, hormonal status) with imaging features within unified predictive models may further refine risk stratification. Advances in multimodal LLMs trained end-to-end on paired imaging and histopathology data hold promise for enhancing image interpretation, but require rigorous prospective, multi-center validation before clinical deployment (23). Finally, implementation studies assessing the impact of nomogram use on biopsy rates, workflow efficiency, and patient outcomes will be crucial to translate these findings into clinical practice.

Nevertheless, certain limitations warrant consideration. First, although the nomograms were externally validated across three centers, prospective evaluation in different geographic regions and imaging platforms is necessary to confirm generalizability. Second, although fully automated deep learning–based lesion delineation could streamline workflows and reduce operator dependence, we intentionally relied on manual expert segmentation with high inter-observer agreement (ICC > 0.85). Prior studies have also shown that AI models trained on manually segmentation data outperformed those based on automatic segmentation when distinguishing malignant from benign breast lesions on ultrasound images (24, 25).

Given the sensitivity of morphometric feature extraction, expert manual delineation remains the reference standard for ensuring anatomical precision. Moreover, additional research is still needed to improve automated breast lesion segmentation on ultrasound to a level that matches the accuracy and reliability of manual expert annotation. By prioritizing high-quality manual segmentation, we ensured that the resulting morphometric features were clinically robust and reproducible. Third, regarding the LLM evaluation, without access to the models' internal attention maps or feature activation layers, as its inherent nature, we cannot definitively determine the extent to which their outputs were driven by genuine visual feature extraction versus reliance on statistical priors or hallucinations expressed through confident language. Fourth, although radiomics derived texture features and patient level clinical variables may provide additional discriminatory value, their use typically depends on standardized image acquisition, specialized post processing, and broad clinical availability. These requirements are not consistently met across many real-world clinical settings. For this reason, our study focused on conventional ultrasound descriptors and BI RADS based morphological characteristics. This approach enabled us to evaluate the intrinsic diagnostic contribution of routinely available and consistently documented imaging biomarkers, and to isolate the value of ultrasound morphology and BI RADS descriptors without introducing variability from heterogeneous or incomplete clinical metadata. Future studies with larger and more standardized datasets will be needed to investigate the added benefit of integrating radiomics, clinical information, and other multimodal data sources. Finally, although the nomograms demonstrated excellent discrimination, calibration performance and decision-curve analyses should be prospectively assessed to confirm net clinical benefit and threshold optimization in practice.

## 5. Conclusion

This comprehensive, multi-center study demonstrates that nomogram models integrating BI-RADS features with quantitative morphometric characteristics deliver superior diagnostic accuracy for breast lesion evaluation, matching or exceeding the performance of expert radiologists and significantly outperforming current LLM interpretations. The fused nomogram offers a user-friendly, interpretable tool for personalized biopsy recommendations and malignancy risk estimation, with the potential to reduce unnecessary interventions and optimize clinical decision-making in breast imaging. Continued efforts to automate segmentation, incorporate additional clinical variables, and validate these models in prospective settings will be essential to fully realize their clinical utility.

**Fig. 1.** Flowchart of patients' selection according to each criterion.

**Fig. 2.** Comprehensive study workflow diagram. The research process began with the collection of 1,747 breast lesion ultrasound images from three centers (two in Iran, one in Turkey). Lesion images underwent manual segmentation to extract 26 quantitative morphometric features (green panel), while radiologists concurrently extracted 10 standard qualitative BI-RADS features (blue panel). These features were used to develop integrated AI nomograms. The diagnostic performance of these nomograms was then systematically compared against three radiologists and two Large Language Models (ChatGPT) across two distinct clinical tasks: (1) Biopsy recommendation (distinguishing BI-RADS <4 from ≥4), and (2) Malignancy prediction for biopsy-eligible lesions.

**Fig. 3.** A three-step feature selection method used to define the best features in identifying breast lesions candidates for biopsy in this study: (a) ICC heat map of morphological features to define contour-independent features; (b) correlation matrix of features to exclude highly correlated features; (c) correlation matrix of independent features; (d) LASSO results applied to independent features.

**Fig. 4.** ROC Curves of nomograms, radiologists, and ChatGPTs in identifying breast lesion candidates for biopsy in the internal validation (a), external validation 1 (b), external validation 2 datasets (c), and all cohorts (d).

**Fig. 5.** A three-step feature selection method used to define the best features in diagnosing benign and malignant breast lesions candidates for biopsy in this study: (a) ICC heat map of morphological features to define contour-independent features; (b) correlation matrix of features to exclude highly correlated features; (c) correlation matrix of independent features; (d) LASSO results applied to independent features.

**Fig. 6.** ROC Curves of nomograms, radiologists, and ChatGPTs in diagnosing benign and malignant breast lesion candidates for biopsy in the internal validation (a), external validation 1 (b), external validation 2 datasets (c), and all cohorts (d).

**Fig. 7.** Fused nomogram models for (a) identifying breast lesion candidates for biopsy, and (b) diagnosing benign and malignant breast lesion candidates for biopsy.

**Fig. 8.** Results of LLMs, three radiologists, and nomograms in the malignancy prediction of breast lesions. (a) A malignant lesion (BI-RADS 4B) that shows a homogeneous background echotexture with fibroglandular tissue composition. The mass has an irregular shape with a non-parallel orientation. The margin is not circumscribed and presents microlobulated edges. The echo pattern is hypoechoic, and no posterior features are observed. No calcifications, architectural distortion, clustered microcysts, or complicated cysts were detected. (b) A malignant lesion (BI-RADS 5) that displays homogeneous background echotexture with fibroglandular tissue composition. The mass is irregularly shaped and non-parallel in orientation. Its margin is not circumscribed, spiculated in nature, with a hypoechoic echo pattern and shadowing as the posterior feature. There are no associated calcifications, architectural distortions, clustered microcysts, or complicated cysts. (c) A malignant lesion (BI-RADS 5) that shows a homogeneous background echotexture with fatty tissue composition. The mass is irregularly shaped and oriented non-parallel. The margin is not circumscribed with microlobulated edges, presenting a hypoechoic echo pattern and lacking posterior features. There are no calcifications, architectural distortions, clustered microcysts, or complicated cysts. (d) A malignant mass (BI-RADS 4B) that also exhibits a homogeneous fibroglandular background echotexture. The mass is irregular in shape, non-parallel in orientation, with microlobulated, not circumscribed margins. The echo pattern is hypoechoic with no posterior features observed. No calcifications, architectural distortion, clustered microcysts, or complicated cysts were identified. (e) A benign lesion (BI-RADS 4A) which demonstrates a homogeneous background echotexture with fatty tissue composition. The mass is irregular in shape and not parallel in orientation, with microlobulated, not circumscribed margins. The echo pattern is complex, cystic and solid, and there are no posterior features present. No calcifications, architectural distortion, or complicated cysts are noted; however, clustered microcysts are observed. (f) A malignant lesion (BI-RADS 5) with homogeneous fibroglandular background echotexture. The mass is oval in shape and parallel in orientation. It has microlobulated, not circumscribed margins with a hypoechoic echo pattern and no posterior features. There are no calcifications, architectural distortions, clustered microcysts, or complicated cysts.

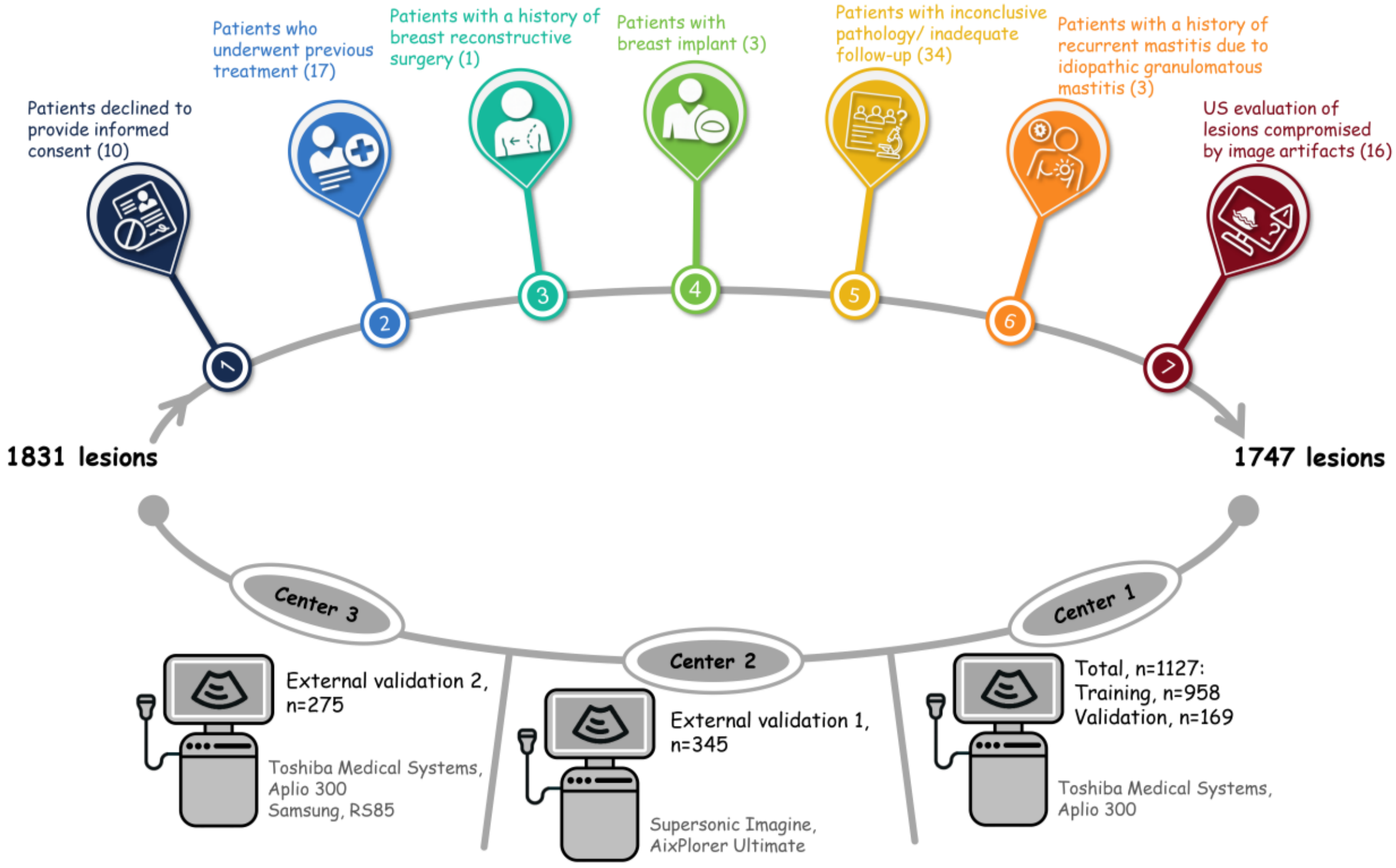

**Fig. 1.** Flowchart of patients' selection according to each criterion.

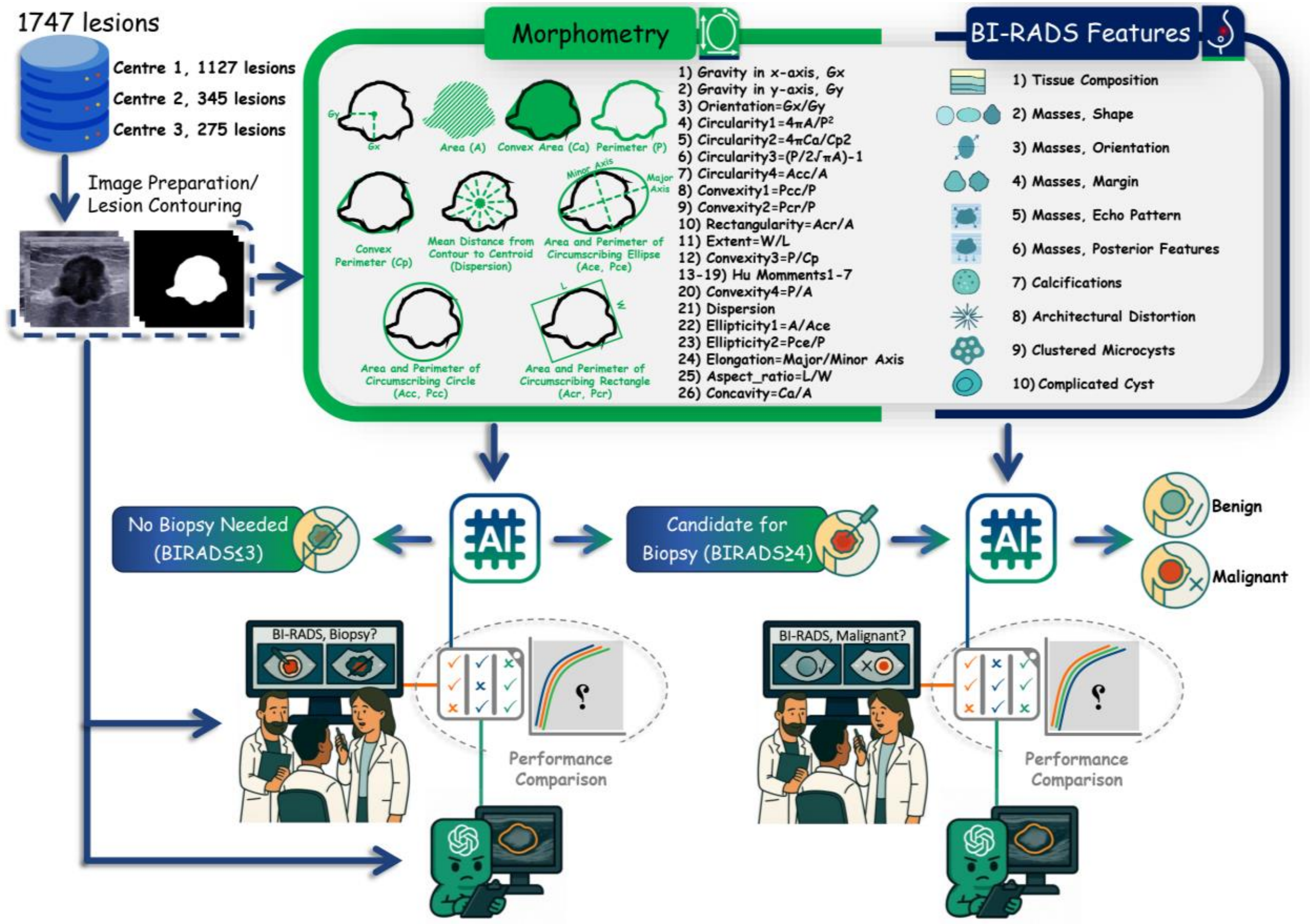

**Fig. 2.** Comprehensive study workflow diagram. The research process began with the collection of 1,747 breast lesion ultrasound images from three centers (two in Iran, one in Turkey). Lesion images underwent manual segmentation to extract 26 quantitative morphometric features (green panel), while radiologists concurrently extracted 10 standard qualitative BI-RADS features (blue panel). These features were used to develop integrated AI nomograms. The diagnostic performance of these nomograms was then systematically compared against three radiologists and two Large Language Models (ChatGPT) across two distinct clinical tasks: (1) Biopsy recommendation (distinguishing BI-RADS <4 from ≥4), and (2) Malignancy prediction for biopsy-eligible lesions.

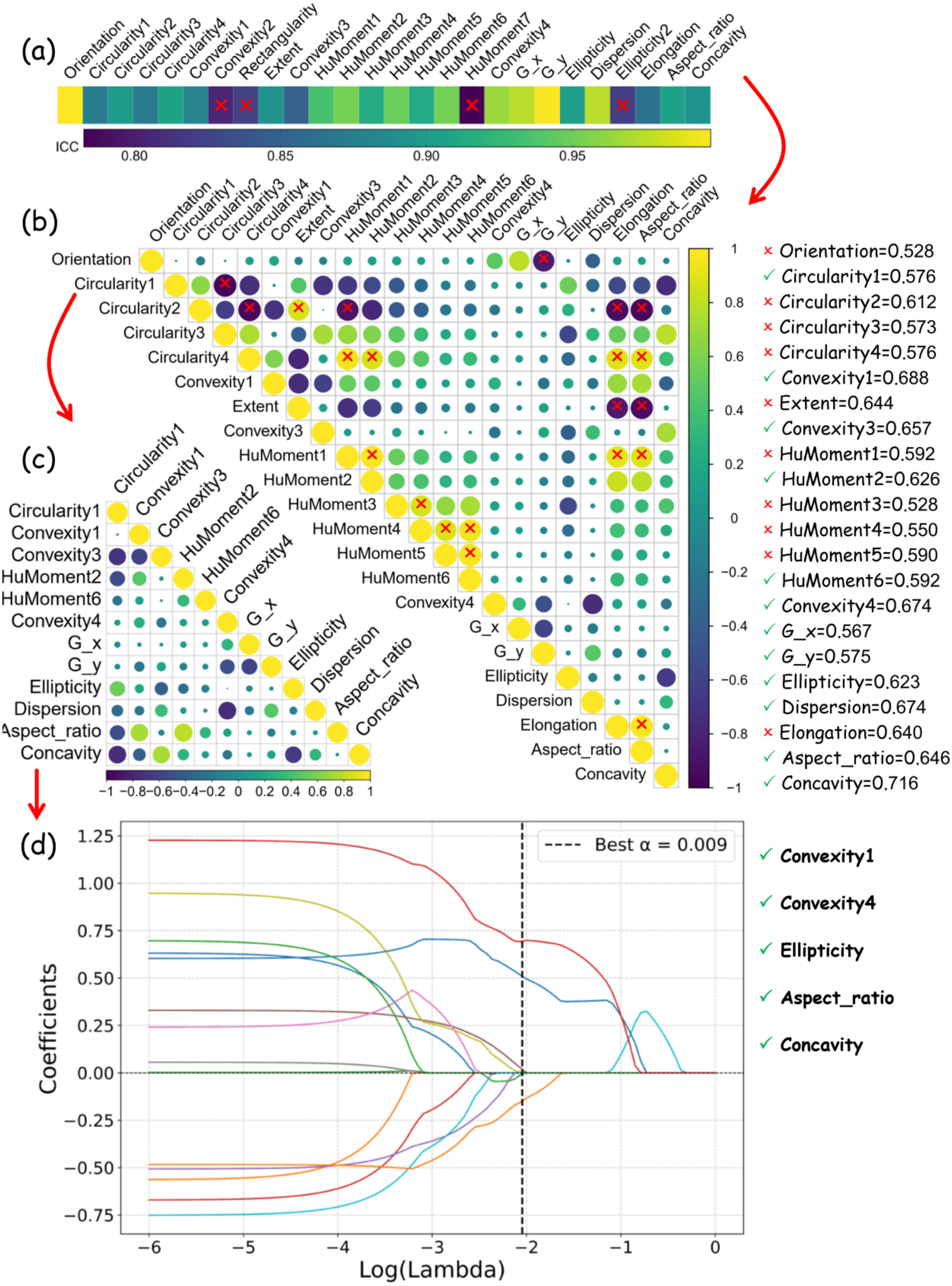

**Fig. 3.** A three-step feature selection method used to define the best features in identifying breast lesions candidates for biopsy in this study: (a) ICC heat map of morphological features to define contour-independent features; (b) correlation matrix of features to exclude highly correlated features; (c) correlation matrix of independent features; (d) LASSO results applied to independent features.

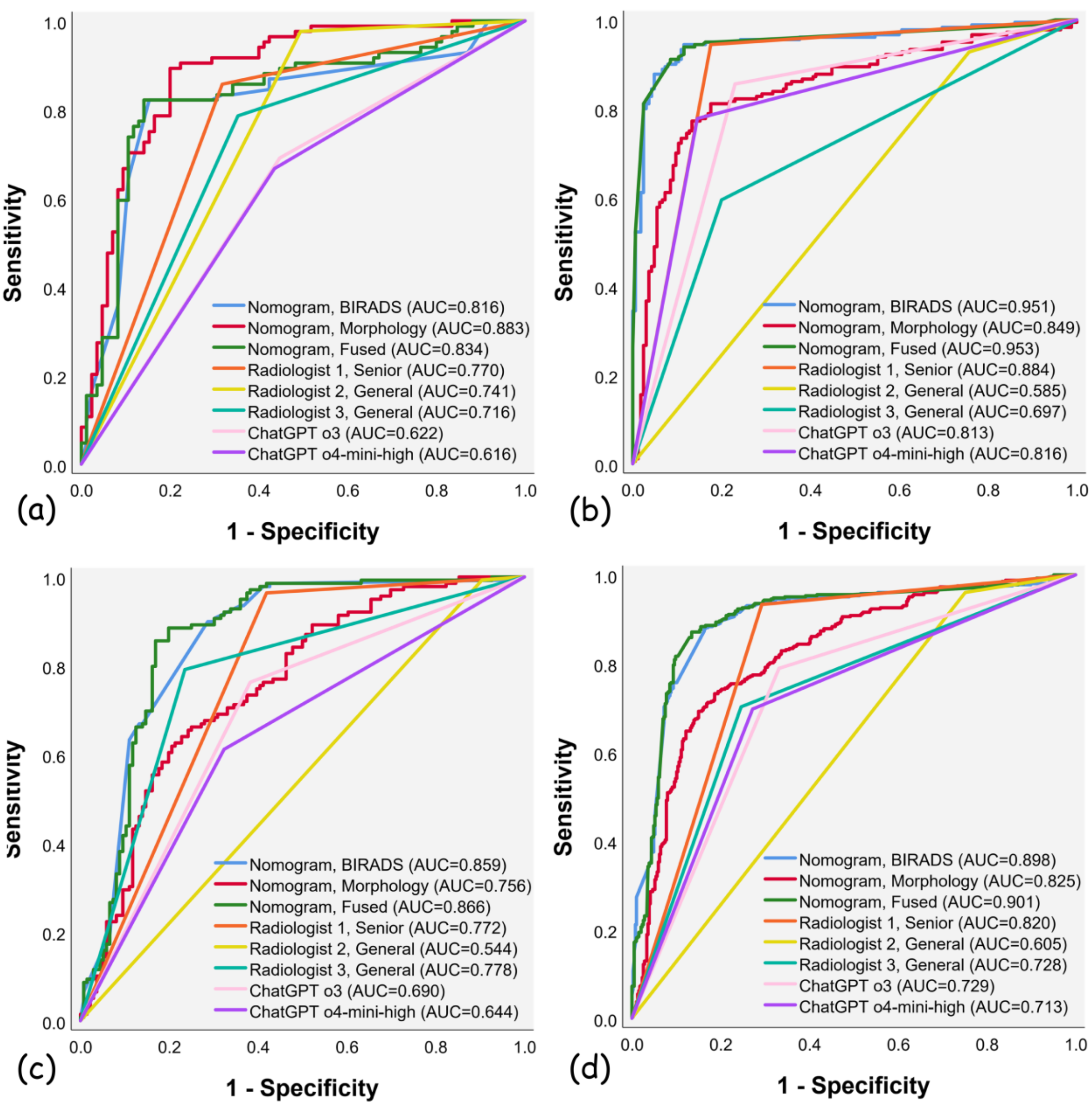

**Fig. 4.** ROC Curves of nomograms, radiologists, and ChatGPTs in identifying breast lesion candidates for biopsy in the internal validation (a), external validation 1 (b), external validation 2 datasets (c), and all cohorts (d).

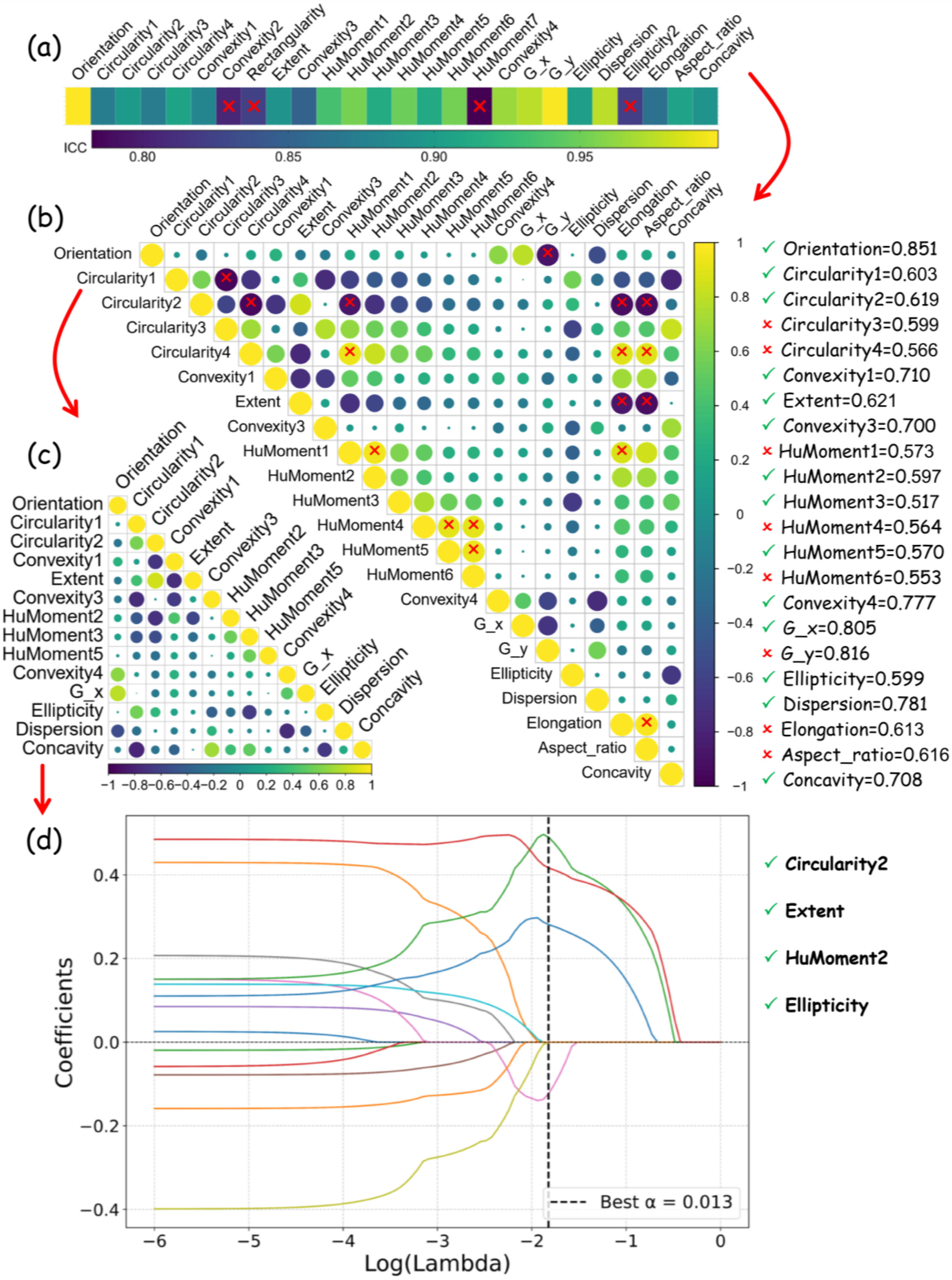

**Fig. 5.** A three-step feature selection method used to define the best features in diagnosing benign and malignant breast lesions candidates for biopsy in this study: (a) ICC heat map of morphological features to define contour-independent features; (b) correlation matrix of features to exclude highly correlated features; (c) correlation matrix of independent features; (d) LASSO results applied to independent features.

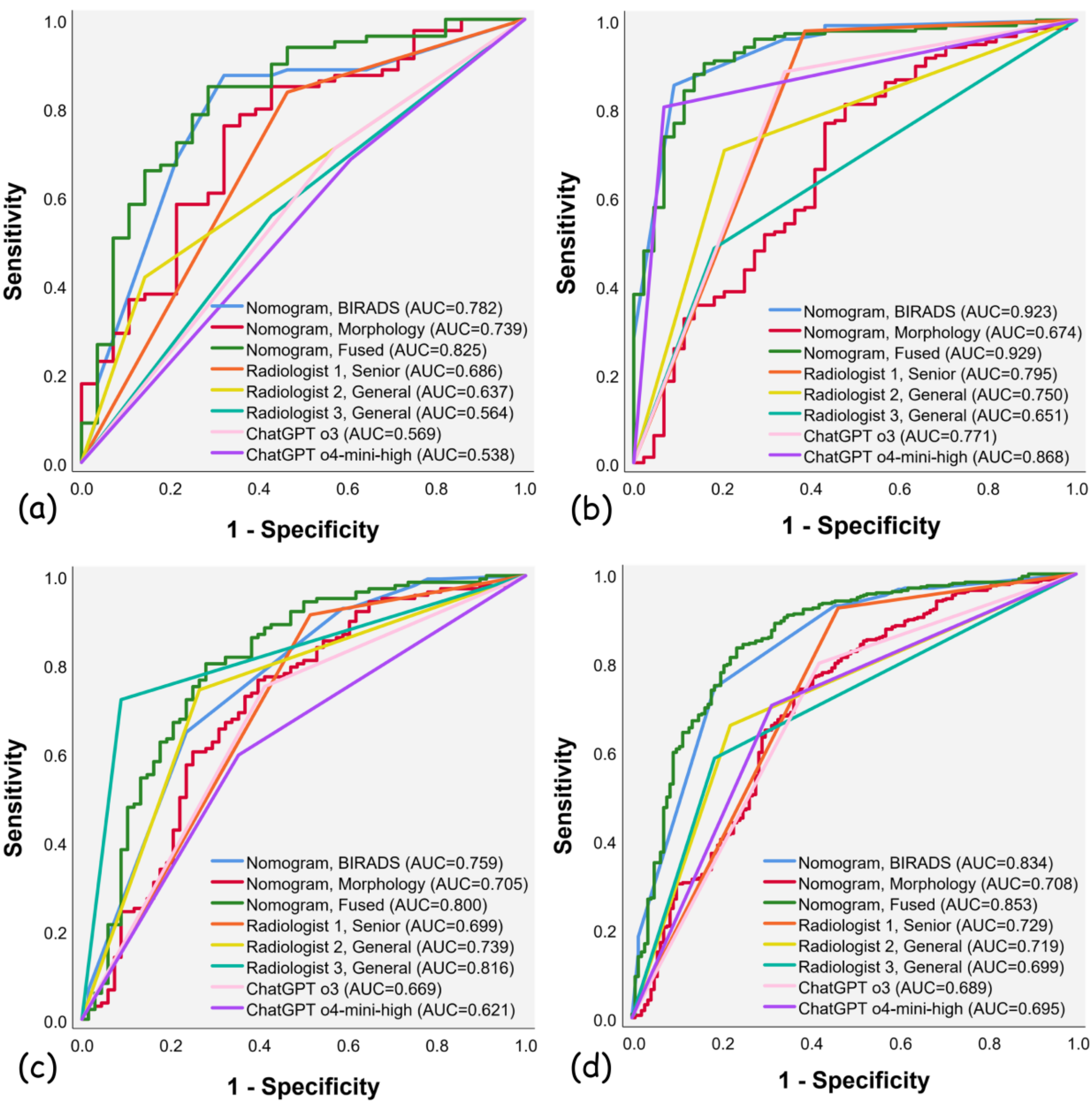

**Fig. 6.** ROC Curves of nomograms, radiologists, and ChatGPTs in diagnosing benign and malignant breast lesion candidates for biopsy in the internal validation (a), external validation 1 (b), external validation 2 datasets (c), and all cohorts (d).

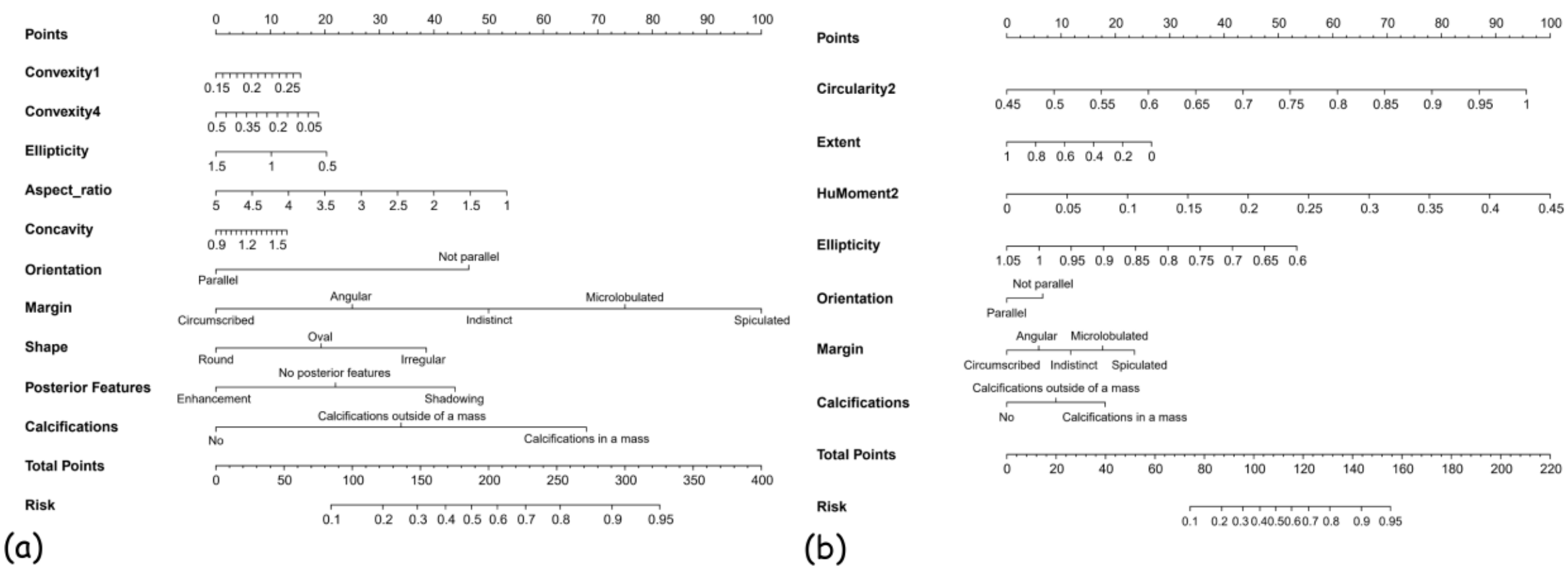

**Fig. 7.** Fused nomogram models for (a) identifying breast lesion candidates for biopsy, and (b) diagnosing benign and malignant breast lesion candidates for biopsy.

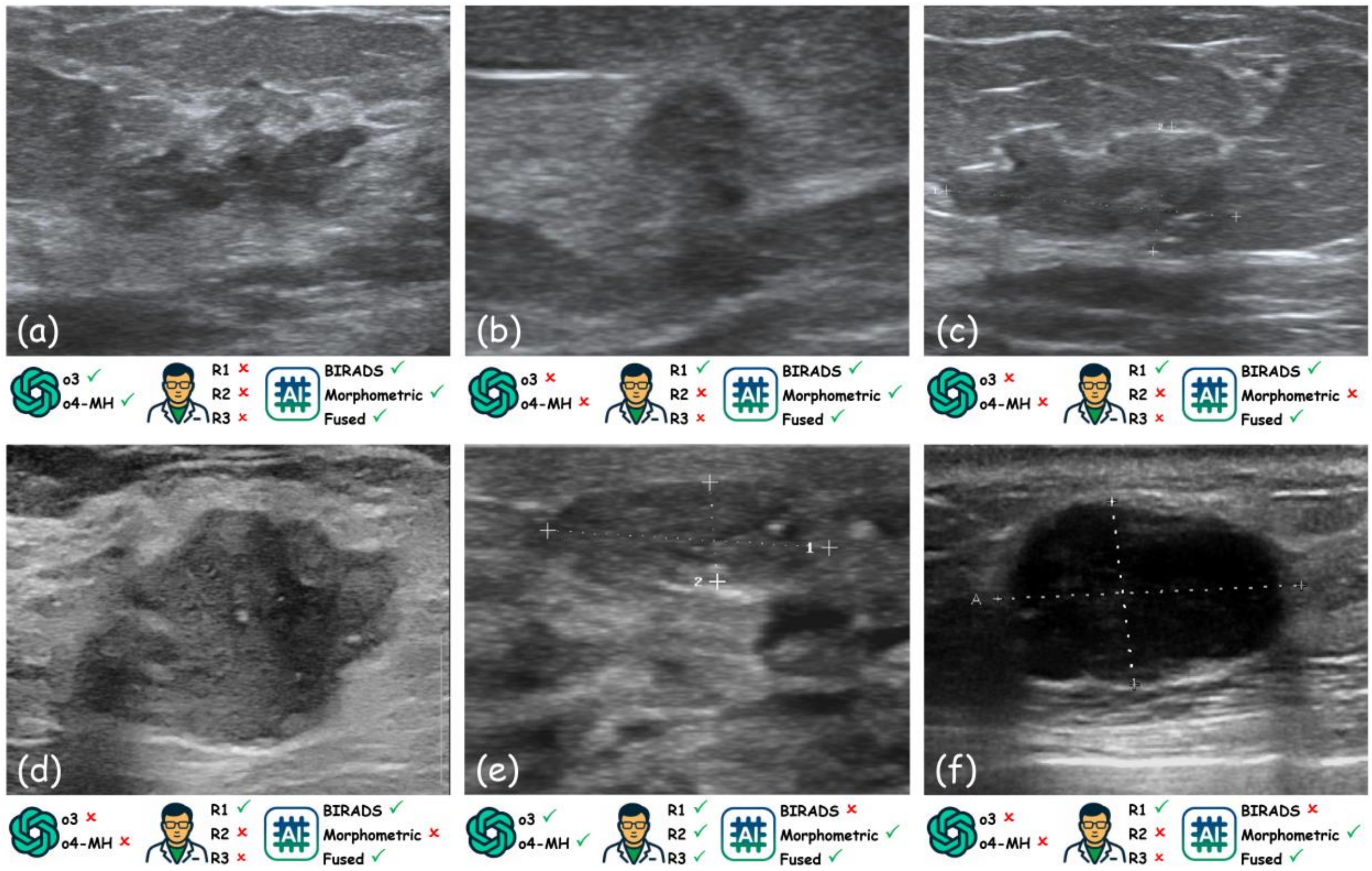

**Fig. 8.** Results of LLMs, three radiologists, and nomograms in the malignancy prediction of breast lesions. (a) A malignant lesion (BI-RADS 4B) that shows a homogeneous background echotexture with fibroglandular tissue composition. The mass has an irregular shape with a non-parallel orientation. The margin is not circumscribed and presents microlobulated edges. The echo pattern is hypoechoic, and no posterior features are observed. No calcifications, architectural distortion, clustered microcysts, or complicated cysts were detected. (b) A malignant lesion (BI-RADS 5) that displays homogeneous background echotexture with fibroglandular tissue composition. The mass is irregularly shaped and non-parallel in orientation. Its margin is not circumscribed, spiculated in nature, with a hypoechoic echo pattern and shadowing as the posterior feature. There are no associated calcifications, architectural distortions, clustered microcysts, or complicated cysts. (c) A malignant lesion (BI-RADS 5) that shows a homogeneous background echotexture with fatty tissue composition. The mass is irregularly shaped and oriented non-parallel. The margin is not circumscribed with microlobulated edges, presenting a hypoechoic echo pattern and lacking posterior features. There are no calcifications, architectural distortions, clustered microcysts,

or complicated cysts. (d) A malignant mass (BI-RADS 4B) that also exhibits a homogeneous fibroglandular background echotexture. The mass is irregular in shape, non-parallel in orientation, with microlobulated, not circumscribed margins. The echo pattern is hypoechoic with no posterior features observed. No calcifications, architectural distortion, clustered microcysts, or complicated cysts were identified. (e) A benign lesion (BI-RADS 4A) which demonstrates a homogeneous background echotexture with fatty tissue composition. The mass is irregular in shape and not parallel in orientation, with microlobulated, not circumscribed margins. The echo pattern is complex, cystic and solid, and there are no posterior features present. No calcifications, architectural distortion, or complicated cysts are noted; however, clustered microcysts are observed. (f) A malignant lesion (BI-RADS 5) with homogeneous fibroglandular background echotexture. The mass is oval in shape and parallel in orientation. It has microlobulated, not circumscribed margins with a hypoechoic echo pattern and no posterior features. There are no calcifications, architectural distortions, clustered microcysts, or complicated cysts.

**Table 1.** Distribution of BI-RADS features and results of univariate and multivariate analyses among breast lesion candidates and non-candidates for biopsy in the training dataset.

| US Findings | Not Candid | Candid | Univariate analysis | | Multivariate analysis | |
|---|---|---|---|---|---|---|
| Tissue composition | | | **Odds Ratio (95%CI)** | **P-value** | **Odds Ratio (95%CI)** | **P-value** |
| Homogeneous background echotexture – fat | 169 | 249 | | | | |
| Homogeneous background echotexture – fibroglandular | 159 | 233 | 0.648 (0.534, 0.787) | <0.001 | - | 0.103 |
| Heterogeneous background echotexture | 23 | 125 | | | | |
| Shape | | | | | | |
| Irregular | 54 | 519 | | | | |
| Oval | 244 | 54 | 0.102 (0.076, 0.136) | <0.001 | 0.575 (405, 0.817) | 0.002 |
| Round | 53 | 34 | | | | |
| Orientation | | | | | | |
| Parallel | 285 | 111 | | | | |
| Not parallel | 66 | 496 | 0.052 (0.037, 0.073) | <0.001 | 0.398 (0.247, 0.640) | <0.001 |
| Margin | | | | | | |
| Circumscribed | 293 | 77 | | | | |
| Not Circumscribed (Angular) | 6 | 27 | | | | |
| Not Circumscribed (Indistinct) | 11 | 38 | | | | |
| Not Circumscribed (Microlobulated) | 33 | 299 | 3.159 (2.781, 3.589) | <0.001 | 2.024 (1.681, 2.432) | <0.001 |
| Not Circumscribed (Spiculated) | 8 | 166 | | | | |
| Echo pattern | | | | | | |
| Anechoic | 48 | 9 | | | | |
| Complex cystic and solid | 3 | 34 | | | | |
| Heterogeneous | 16 | 95 | | | | |
| Hyperechoic | 4 | 0 | - | 0.066 | - | - |
| Hypoechoic | 280 | 461 | | | | |
| Isoechoic | 0 | 8 | | | | |
| Posterior features | | | | | | |
| Enhancement | 50 | 18 | | | | |
| No posterior features | 287 | 457 | 5.130 (3.502, 7.513) | <0.001 | 2.142 (1.179, 3.889) | 0.012 |
| Shadowing | 14 | 132 | | | | |
| Calcifications | | | | | | |
| Calcifications in a mass | 6 | 107 | | | | |
| Calcifications outside of a mass | 0 | 8 | 0.265 (0.172, 0.407) | <0.001 | 0.607 (0.386, 1.001) | 0.042 |
| No | 345 | 492 | | | | |
| Architectural distortion | | | | | | |
| No | 351 | 607 | - | - | - | - |
| Yes | 0 | 0 | | | | |
| Clustered microcysts | | | | | | |
| No | 347 | 601 | - | 1.000 | - | - |
| Yes | 4 | 6 | | | | |
| Complicated cyst | | | | | | |
| No | 348 | 598 | - | 0.551 | - | - |
| Yes | 3 | 9 | | | | |

**Table 2.** Diagnostic performance of nomograms, radiologists, and ChatGPTs in identifying breast lesion candidates for biopsy in all cohorts.

| **Method** | **Predicted Class** | **True Class** | | **MCC** | **Sen (%)** | **Spec (%)** | **Acc (%)** | **AUC (95% CI)** |
|---|---|---|---|---|---|---|---|---|
| | | **Benign** | **Malignant** | | | | | |
| Nomogram (BIRADS) | Not Candid | 277 | 27 | 0.668 | 93.3 | 71.7 | 82.7 | 0.898 (0.874, 0.922) |
| | Candid | 109 | 376 | | | | | |
| Nomogram (Morphology) | Not Candid | 288 | 97 | 0.505 | 75.9 | 74.6 | 75.3 | 0.825 (0.795, 0.854) |
| | Candid | 98 | 306 | | | | | |
| Nomogram (Fused) | Not Candid | 278 | 26 | 0.674 | 93.5 | 72.0 | 83.0 | 0.901 (0.877, 0.925) |
| | Candid | 108 | 377 | | | | | |
| Radiologist 1 (Senior) | Not Candid | 273 | 27 | 0.659 | 93.3 | 70.7 | 82.2 | 0.820 (0.789, 0.851) |
| | Candid | 113 | 376 | | | | | |
| Radiologist 2 (General) | Not Candid | 96 | 16 | 0.299 | 96.0 | 24.9 | 61.2 | 0.605 (0.565, 0.644) |
| | Candid | 290 | 387 | | | | | |
| Radiologist 3 (General) | Not Candid | 291 | 120 | 0.456 | 70.2 | 75.4 | 72.5 | 0.728 (0.692, 0.764) |
| | Candid | 95 | 283 | | | | | |
| ChatGPT o3 | Not Candid | 258 | 86 | 0.461 | 78.9 | 66.8 | 73.0 | 0.729 (0.693, 0.765) |
| | Candid | 128 | 318 | | | | | |
| ChatGPT o4-mini-high | Not Candid | 281 | 122 | 0.425 | 69.7 | 72.8 | 71.2 | 0.713 (0.676, 0.749) |
| | Candid | 105 | 281 | | | | | |

**Table 3.** Distribution of BI-RADS features and results of univariate and multivariate analyses among benign and malignant breast lesion candidates for biopsy in the training dataset.

| US Findings | Benign | Malignant | Univariate analysis | | Multivariate analysis | |
|---|---|---|---|---|---|---|
| Tissue composition | | | **Odds Ratio (95%CI)** | **P-value** | **Odds Ratio (95%CI)** | **P-value** |
| Homogeneous background echotexture – fat | 75 | 174 | | | | |
| Homogeneous background echotexture – fibroglandular | 65 | 168 | 0.759 (0.592, 0.972) | 0.029 | - | 0.201 |
| Heterogeneous background echotexture | 19 | 106 | | | | |
| Shape | | | | | | |
| Irregular | 121 | 398 | | | | |
| Oval | 27 | 27 | 0.622 (0.453, 0.853) | 0.003 | - | 0.113 |
| Round | 11 | 23 | | | | |
| Orientation | | | | | | |
| Parallel | 53 | 58 | 0.297 (0.194, 0.457) | <0.001 | 0.435 (0.247, 0.767) | 0.004 |
| Not parallel | 106 | 390 | | | | |
| Margin | | | | | | |
| Circumscribed | 28 | 49 | | | | |
| Not Circumscribed (Angular) | 21 | 6 | | | | |
| Not Circumscribed (Indistinct) | 17 | 21 | 1.481 (1.291, 1.699) | <0.001 | 1.454 (1.209, 1.754) | 0.003 |
| Not Circumscribed (Microlobulated) | 70 | 229 | | | | |
| Not Circumscribed (Spiculated) | 21 | 143 | | | | |
| Echo pattern | | | | | | |
| Anechoic | 4 | 5 | | | | |
| Complex cystic and solid | 1 | 33 | | | | |
| Heterogeneous | 23 | 72 | - | 0.100 | - | - |
| Hyperechoic | 0 | 0 | | | | |
| Hypoechoic | 131 | 330 | | | | |
| Isoechoic | 0 | 8 | | | | |
| Posterior features | | | | | | |
| Enhancement | 9 | 9 | | | | |
| No posterior features | 122 | 335 | 1.577 (1.044, 2.381) | 0.030 | - | 0.731 |
| Shadowing | 28 | 104 | | | | |
| Calcifications | | | | | | |
| Calcifications in a mass | 8 | 99 | | | | |
| Calcifications outside of a mass | 0 | 8 | 0.408 (0.279, 0.597) | <0.001 | 0.591 (0.433, 0.825) | <0.001 |
| No | 151 | 341 | | | | |
| Architectural distortion | | | | | | |
| No | 159 | 448 | - | - | - | - |
| Yes | 0 | 0 | | | | |
| Clustered microcysts | | | | | | |
| No | 156 | 445 | - | 0.188 | - | - |
| Yes | 3 | 3 | | | | |
| Complicated cyst | | | | | | |
| No | 154 | 444 | - | 0.058 | - | - |
| Yes | 5 | 4 | | | | |

**Table 4.** Diagnostic performance of nomograms, radiologists, and ChatGPTs in diagnosing benign and malignant breast lesions candidates for biopsy in all cohorts.

| **Method** | **Predicted Class** | **True Class** | | **MCC** | **Sen (%)** | **Spec (%)** | **Acc (%)** | **AUC (95% CI)** |
|---|---|---|---|---|---|---|---|---|
| | | **Benign** | **Malignant** | | | | | |
| Nomogram (BIRADS) | Benign | 76 | 29 | 0.515 | 92.3 | 54.3 | 82.0 | 0.834 (0.794, 0.875) |
| | Malignant | 64 | 349 | | | | | |
| Nomogram (Morphology) | Benign | 58 | 46 | 0.324 | 87.8 | 41.4 | 75.2 | 0.708 (0.589, 0.703) |
| | Malignant | 82 | 332 | | | | | |
| Nomogram (Fused) | Benign | 88 | 32 | 0.573 | 91.5 | 62.9 | 83.8 | 0.853 (0.814, 0.892) |
| | Malignant | 52 | 346 | | | | | |
| Radiologist 1 (Senior) | Benign | 75 | 29 | 0.509 | 92.3 | 53.6 | 81.8 | 0.729 (0.675, 0.784) |
| | Malignant | 65 | 349 | | | | | |
| Radiologist 2 (General) | Benign | 109 | 129 | 0.390 | 65.9 | 77.9 | 69.1 | 0.719 (0.670, 0.768) |
| | Malignant | 31 | 249 | | | | | |
| Radiologist 3 (General) | Benign | 114 | 157 | 0.355 | 58.5 | 81.4 | 64.7 | 0.699 (0.651, 0.748) |
| | Malignant | 26 | 221 | | | | | |
| ChatGPT o3 | Benign | 81 | 76 | 0.365 | 79.9 | 57.9 | 73.9 | 0.689 (0.635, 0.743) |
| | Malignant | 59 | 302 | | | | | |
| ChatGPT o4-mini-high | Benign | 96 | 112 | 0.353 | 70.4 | 68.6 | 69.9 | 0.695 (0.643, 0.747) |
| | Malignant | 44 | 266 | | | | | |